\documentclass[]{spie}  

\usepackage{amsmath,amsfonts,amssymb}
\usepackage[colorlinks=true, allcolors=blue]{hyperref}
\usepackage{graphicx}
\usepackage{subfigure}

\title{Enhancing the sensitivity of next-generation X-ray imaging detectors with artificial intelligence and advanced event reconstruction algorithms}

\author[a,b]{D.~ R.~Wilkins}
\author[c]{A.~Poliszczuk}
\author[c,d]{A.~Y.~Pan}
\author[c,d]{L.~Sajkov}
\author[c,d]{S.~W.~Allen}
\author[e]{M.~Heine}
\author[e]{C.~E.~Grant}
\author[e]{M.~W.~Bautz}
\author[c]{T.~Chattopadhyay}
\author[f]{K.~Donlon}
\author[c]{S.~Herrmann}
\author[e]{B.~LaMarr}
\author[e]{E.~D.~Miller}
\author[c]{P.~Orel}

\affil[a]{Department of Astronomy, The Ohio State University, 140~W~18th~Ave, Columbus, OH 43210, USA}
\affil[b]{Center for Cosmology and AstroParticle Physics (CCAPP), The Ohio State University, 191~W~Woodruff Ave, Columbus, OH 43210, USA}
\affil[c]{Kavli Institute for Particle Astrophysics and Cosmology, Stanford University, 452~Lomita Mall, Stanford, CA 94305, USA}
\affil[d]{Department of Physics, Stanford University, 382~Via Pueblo Mall, Stanford, CA 94305, USA}
\affil[e]{MIT Kavli Institute for Astrophysics and Space Research, 77~Massachusetts Ave, Cambridge, MA 02139, USA}
\affil[f]{MIT Lincoln Laboratory, 244~Wood Street, Lexington, MA 02421, USA}

\authorinfo{Correspondence e-mail: wilkins.401@osu.edu}

\begin{document} 
\maketitle

\begin{abstract}
Advanced algorithms incorporating artificial intelligence and machine learning (AI/ML) enhance the sensitivity of X-ray imaging detectors and the scientific capabilities of future X-ray missions. In orbit, current X-ray instruments are limited in their sensitivity by (1) the instrumental background, induced by cosmic rays which produce signals that can be confused for genuine, astrophysical X-rays, and (2) the ability to reconstruct the detected photon events, degrading both the quantum efficiency and energy resolution at the lowest energies, where much discovery space resides. We report on the development of prototype algorithms designed to operate on the raw frame-level data to provide improved identification of particle-induced background events and enhanced energy reconstruction. These algorithms consider the contextual information from all signals in a frame holistically, and are built upon physics-motivated models of charge diffusion and signal generation within the detector. Using high fidelity simulations, we show that following recent developments, prototype ML algorithms can achieve a reduction of the unrejected particle background by up to 68 per cent compared with traditional filtering methods when operated in an aggressive mode suitable for source detection in X-ray imaging surveys, or up to 40 per cent in a more conservative mode designed to prioritize accurate measurements of the spectrum. Moreover, we find that next-generation event reconstruction algorithms can provide an improvement in the sensitivity and energy resolution of CCD-like detectors at event energies below 1\,keV, and can provide additional information to aid in the filtering of background events, and to reduce the impact of photon pile-up. We present new laboratory data that demonstrates the performance of the algorithm on the MIT-LL CCID-93 CCD detector. Together with the capabilities of next-generation high-speed, low-noise detectors, these algorithms can satisfy the requirements for future X-ray flagship missions.
 
\end{abstract}


\section{INTRODUCTION}
\label{sec:intro}  
Many of the flagship science cases of next-generation space-based X-ray observatories depend upon the sensitivity of X-ray imaging detectors to low surface brightness X-ray sources at moderate to high redshifts.\cite{athena_science,lynx_science} These include deep X-ray surveys to find the seeds of supermassive black holes in the early Universe, and measurements of the outskirts of galaxy clusters as well as the circumgalactic medium (CGM) of individual galaxies and the warm-hot component of the intergalactic medium (WHIM) that represent a key tracer of structure formation, AGN feedback and the baryon cycle within galactic environments.

New detector technologies are enabling high-frame rate, low-read noise X-ray imaging detectors\cite{xray_speed,stueber+2025}, including the DEPFET sensor array that forms the basis of the \textit{NewAthena Wide Field Imager} (WFI)\cite{wfi}, and next-generation CCD detectors that enable large format, small pixel imaging arrays for a future flagship X-ray observatory, for example those based upon the \textit{Lynx High Definition X-ray Imager} (HDXI) concept\cite{lynx_hdxi}. In addition, novel readout technologies including the SiSeRO readout stage can be integrated into next-generation CCDs or form the basis of an active pixel sensor, enabling high frame rates with less than 0.5 electron RMS read noise\cite{sisero_jatis_1,sisero_jatis_2}.

There are, however, two significant limitations to the sensitivity of these detectors to low surface brightness X-ray sources. On one hand, there is the instrumental background. Over the 2-10\,keV energy band, this is dominated by charged cosmic ray particles interacting with the spacecraft and the detector.\cite{wfi_bkg} Primary cosmic ray particles may either interact directly with the detector, depositing energy into the pixels that induce the background signal, or they may interact elsewhere within the spacecraft or camera assembly, producing secondary particles (including protons, electrons, and, indeed, X-ray photons) that then go on to deposit energy in the detector. Many of these background events can be filtered out of the science data using traditional techniques that identify background events based upon the pattern of charge deposited across the pixels of the detector (\textit{e.g.} the extended tracks produced by charged particles), or the total event energy (as charged particle events often deposit more energy in a single event than be attributed to a single X-ray photon that could have been focused by the mirrors). While these traditional filtering techniques can identify around 98 per cent of background events,\cite{spie_2020} the remaining unrejected events that do not meet the filtering criteria in pixel pattern or event energy (so-called `valid events') produce a significant signal that limits the sensitivity of the detector to the faintest sources.

On the other hand is the ability of the detector to detect the lowest energy photons, and to accurately reconstruct these events to measure their energies. When a photon is absorbed in a silicon detector, a cloud of electrons is liberated. This charge cloud then drifts through an electric field to reach the charge collection gates through which the corresponding signal is read. Measurement of the photon energy requires an accurate measurement of the number of electrons in this charge packet. As the electrons drift, however, the cloud diffuses, and depending where the photon was absorbed relative to the pixel boundaries, some of the electrons will diffuse into the neighboring pixels, producing split events, a phenomenon that is well understood in CCDs and similar pixelated X-ray detectors. The signal from neighboring pixels must be added to obtain the photon energy measurement.\cite{trad_event} In order to limit the impact of electronic noise, only the signal from pixels exceeding the \textit{split threshold} level are summed, meaning that the energy carried by electrons into the outer pixels of the event that do not reach the split threshold can be lost, leading to a source of  error in the photon energy measurement. This effect is particularly pronounced for low energy photons in back-illuminated devices. In a back-illuminated device, low energy photons are absorbed in a relatively shallow depth of silicon, furthest from the charge collection gates, increasing the number of electrons that will diffuse into outer pixels that do not reach the split threshold, leading to a systematic underestimate in the energy, degrading the spectral resolution and reducing the sensitivity of the device to the lowest energy photons that are critical to many sciences cases that require the detection of soft X-ray sources, and X-ray sources at significant redshifts.\cite{miller_charge_diff,spie_2024}

These proceedings report on an ongoing program to to develop enhanced algorithms to mitigate these two limitations. We are developing artificial intelligence and machine learning (AI/ML) algorithms to provide enhanced filtering of the cosmic ray-induced background in X-ray imaging detectors, and next-generation event reconstruction algorithms for X-ray CCD, CMOS, DEPFET and similar detectors to enhance the sensitivity and spectral resolution at low photon energies. These algorithms are built upon a detailed physical modeling of how both charged particles and X-ray photons interact with the detector.

\section{AI/ML FILTERING OF THE INSTRUMENTAL BACKGROUND}
We are developing machine learning algorithms to provide enhanced filtering of the cosmic ray-induced background in X-ray imaging detectors.\cite{spie_2020,spie_2022,spie_2024} Traditional filtering algorithms, based upon the \textit{ASCA} grading scheme\cite{xis}, are employed on all current-generation X-ray CCDs, including \textit{Chandra ACIS}, \textit{XMM-Newton EPIC} and \textit{XRISM Xtend}. These traditional algorithms filter events based upon the pattern of pixels illuminated in the event (single and double pixel events, in addition to triple and quadruple pixel events with valid patterns in small pixel devices, are considered valid X-ray events), and the total event energy (which must be within the range of photon energies that can be focused by the mirrors).

One significant limitation of this traditional filtering method is that each event is considered in isolation. Multiple background events can originate from the shower of secondaries produced by a single primary cosmic ray particle, meaning that additional information is available when events are considered in the context within which they appear within a frame read out from the detector. Machine learning algorithms are able to leverage this contextual information to provide enhanced filtering of the background.\cite{spie_2020,spie_2022}

We have developed a prototype two-stage machine learning algorithm to classify individual events within frames read from next-generation CCD and DEPFET detectors, based upon a convolutional neural network (CNN) architecture.\cite{poliszczuk+2024} The algorithm operates on the raw pixel-by-pixel values read out from the detector in a single frame and considers all of the events appearing within a frame simultaneously. The first stage of the algorithm employs a convolutional neural network that classifies the entire frame as containing (1) only genuine astrophysical X-ray events, (2) only cosmic ray-induced events or (3) a mixture of X-ray and cosmic ray events.

In the latest version of the algorithm, the CNN stage is based upon the \textsc{ResNet} architecture\cite{resnet}. Rather than simply using consecutive layers of learned convolutional filters to learn arbitrary features within the input image, \textsc{ResNet} employs `residual blocks,' reformulating each layer as a residual function that is learned and added to that layer's input. In addition, the \textsc{ResNet} architecture has skip connections that bypass intermediate layers of the neural network if they add no value. Combined, these aspects of the architecture allow for the training of deeper networks with improved accuracy and faster convergence. The ability to train deep networks is important in image recognition, ensuring that our algorithm is able to effectively learn the features that distinguish background events from genuine astrophysical X-rays.

\begin{figure}
    \centering
    \includegraphics[width=0.9\linewidth]{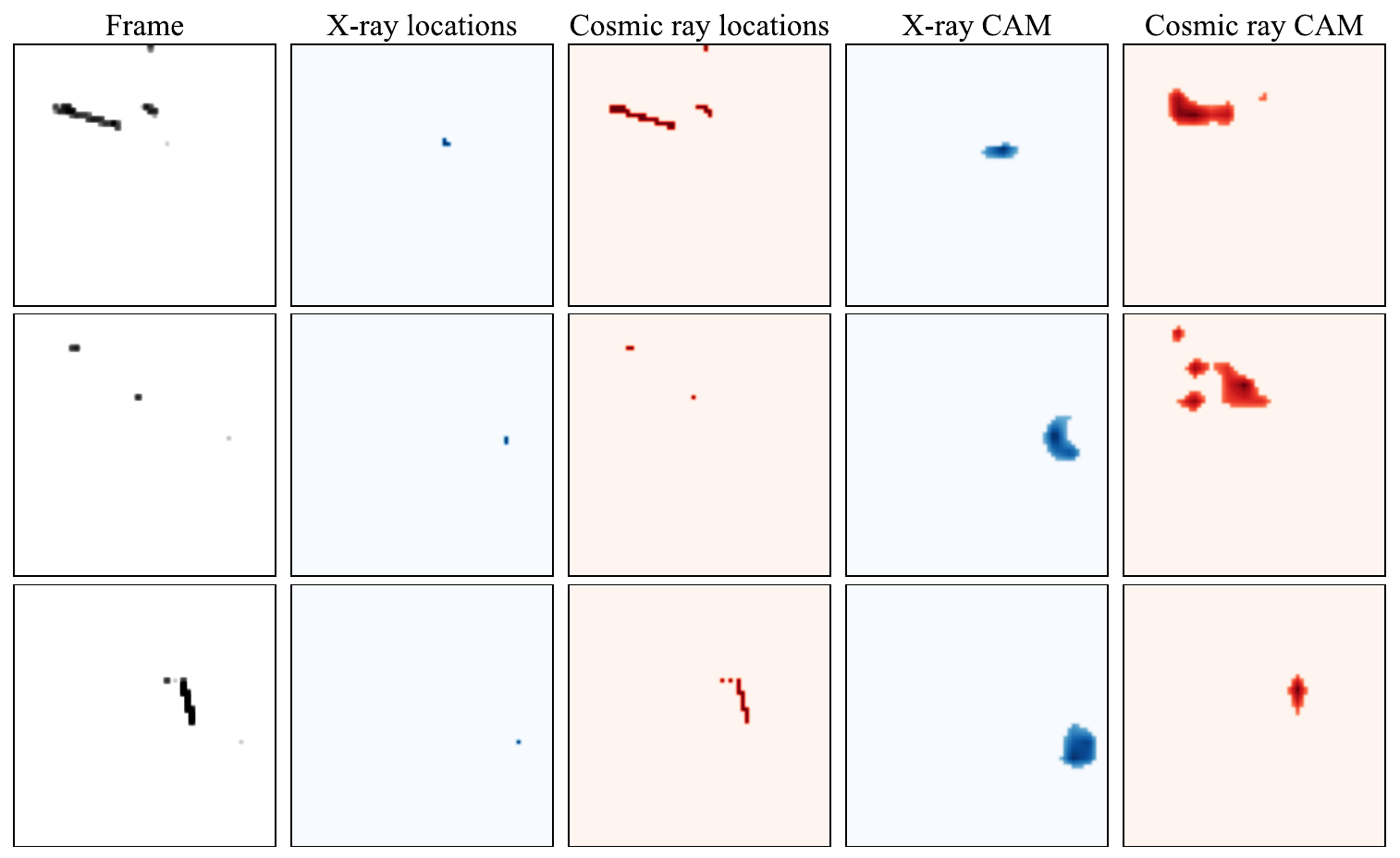}
    \caption{Class activation maps (CAMs) showing the pixels identified by the convolutional neural network (CNN) algorithm as containing either astrophysical X-ray or cosmic ray events. The left column shows the input frame image, with shading corresponding to the recorded pixel values. The second and third columns show the true locations of pixels illuminated by X-ray and cosmic ray events, and the fourth and fifth columns show the CAMs for the X-ray and cosmic ray identifications, respectively.}
    \label{fig:cams}
\end{figure}

The neural network yields a single classification value for each input frame. Therefore, in order to identify X-ray and background events within the image frame, our two-stage algorithm utilizes \textit{class activation maps} (CAMs)\cite{cam}. These are computed from the final convolutional layer of the neural network and identify the pixels within the frame that cause the network to assign a specific classification. It is these CAMs that identify pixels as containing either X-ray or cosmic ray events (Figure~\ref{fig:cams}). The event itself is reconstructed using the traditional \textit{ASCA} scheme, and the CAM values are input into a random forest (RF) classification algorithm\cite{random_forest} along with the total event energy to yield the final classification of each event. This approach to \textit{weakly supervised object localization} (WSOL) overcomes problems encountered training pixel-by-pixel classification and semantic segmentation algorithms on sparse image data and allows for efficient training of the algorithm on image frames from high frame rate detectors in which many of the pixels contain no signal.\cite{poliszczuk+2024}

The final classification of each event is a number between 0 and 1 and can be interpreted as the probability that the event is cosmic ray-induced background. This classification can be added as a column to the final event list used for science analysis, and the end user can choose to filter the events based upon a background probability threshold that can be tuned to a specific science case, trading off between completeness (minimal rejection of genuine X-rays) and purity (maximal rejection of background).

While the prototype AI/ML filtering algorithms run on conventional PC platforms in the ground segment, their operation requires all of the event data (\textit{i.e.} the raw pixel values from every pixel in which a signal is detected above the electronic readout noise level) from every frame read from the detector. Such a model may not be feasible on missions for which telemetry is limited. Future development will therefore include the compression of these algorithms to a form that can be run on lightweight compute hardware, including FPGAs, on board the spacecraft, receiving the raw pixel-by-pixel data stream from the detector electronics in an `edge-AI' implementation to process events in real time.

\subsection{Training data for background filtering algorithms}
The algorithms are trained upon high fidelity simulations of interactions between cosmic ray protons and the detector using \textsc{geant4}\cite{geant4} to provide as realistic event data as possible for next-generation X-ray imaging detectors. For the purposes of developing and testing the algorithm, we employ simulations specifically developed for the \textit{NewAthena Wide Field Imager}, in which cosmic ray protons interact with a realistic mass model of the camera head and DEPFET detector, depositing energy within the 3D detector volume which then goes on to produce the background signals and events that are detected.\cite{grant+2020,miller_jatis,heine+2026} The full series of interactions, including any secondary particles or photons that are produced, are followed from every primary cosmic ray particle to produce realistic sets of correlated events that would arise from a single shower of secondaries.

In addition to cosmic ray-induced signals, simulated astrophysical X-ray events are added to each frame, depositing their energy at a single location within the silicon at a depth drawn from an exponential distribution applicable to the attenuation depth specific to the energy of that photon. Once energy is deposited in the silicon, both from X-ray and charged particle events, a semi-analytic model is applied to simulate the diffusion of the resulting charge cloud as the electrons drift and diffuse towards the charge collection gates, predicting the recorded signal in each pixel\cite{poliszczuk+2023}. This model is calibrated against full physical simulations of electron drift and diffusion in CCDs\cite{miller+2022,lamarr+2022} using the \textsc{poisson\_ccd} code\cite{poissonccd}, but allows rapid simulation of large numbers of events. An overview of the event simulation procedure is presented in Ref.~\citenum{spie_2024}.

The training set comprised 500,000 simulated frames, based upon a \textit{NewAthena WFI} detector model, with a mean rate of one cosmic ray (including all of the secondary events in the shower resulting from a single primary proton) and one astrophysical X-ray per frame, corresponding to the approximate cosmic ray and X-ray event rates when observing a faint X-ray source.\cite{grant+2020,miller_jatis}. Cosmic ray events in the training set were drawn from a library of 150,872 events simulated using \textsc{geant4}. We require that all of the frames in the training set have at least one cosmic ray shower and one `invalid' event with energy greater than 20\,keV somewhere within the frame. 75,872 of the simulated cosmic ray showers produce `valid' events that accompany the invalid event within the frame, and an additional 75,000 showers produce only invalid events. The validation set upon which the performance of the algorithm was tested comprised a further 10,000 frames, built from a separate library of 19,062 simulated cosmic ray showers that were not included in the training set, 9,562 of which produced `valid' events and 9,500 produced only invalid events.

\subsection{Photon energy information in the training data}
Two sets of astrophysical photon events were used in the training data: one in which the photons follow a power law spectrum with a photon index of $\Gamma = 1$, to emulate astrophysical X-ray sources such as AGN. A second set was simulated in which the astrophysical photons have the same energy spectrum as the valid background events (\textit{i.e.} those not removed by traditional filtering methods). We use these two training sets to understand the degree to which classification of events is determined by the energy spectrum of individual events, and the degree to which the algorithm learns the pixel patterns and the context of within which background events appear.

We find that when the algorithm is trained upon simulated astrophysical X-ray and background events with distinct spectra, the classifier is able to use just the measured energy of an isolated `valid' event to probabilistically classify that event as a genuine X-ray or a cosmic ray-induced background event (\textit{i.e.} learning that a higher energy event is more likely to be cosmic ray-induced) even if no further contextual information is available within the frame. While this training increases the rate at which background events are identified, the algorithm is more likely to falsely identify higher energy photons as background, distorting the spectrum that remains after filtering.

Training the algorithm using a photon energy distribution matching the background event spectrum removes this energy information from the training data, forcing the algorithm to learn only the pixel patterns and the context of events within a frame. In the latter case, the true spectrum of the unfiltered astrophysical photons is preserved. The former can be used to train a more aggressive filtering algorithm that is suitable for the detection of faint sources within deep X-ray surveys, at the expense of precise spectral calibration. The latter can be used to train a more conservative mode of the algorithm in which the background is not filtered to the same degree, but no energy dependence is introduced in the false-positive rate, preserving the shape of the measured astrophysical X-ray spectrum to enable detailed spectral analysis.

\subsection{Performance of prototype background filtering algorithm}

\begin{figure}
    \centering
    \subfigure[] {
    \label{fig:ai_performance:roc}
    \includegraphics[width=0.45\linewidth]{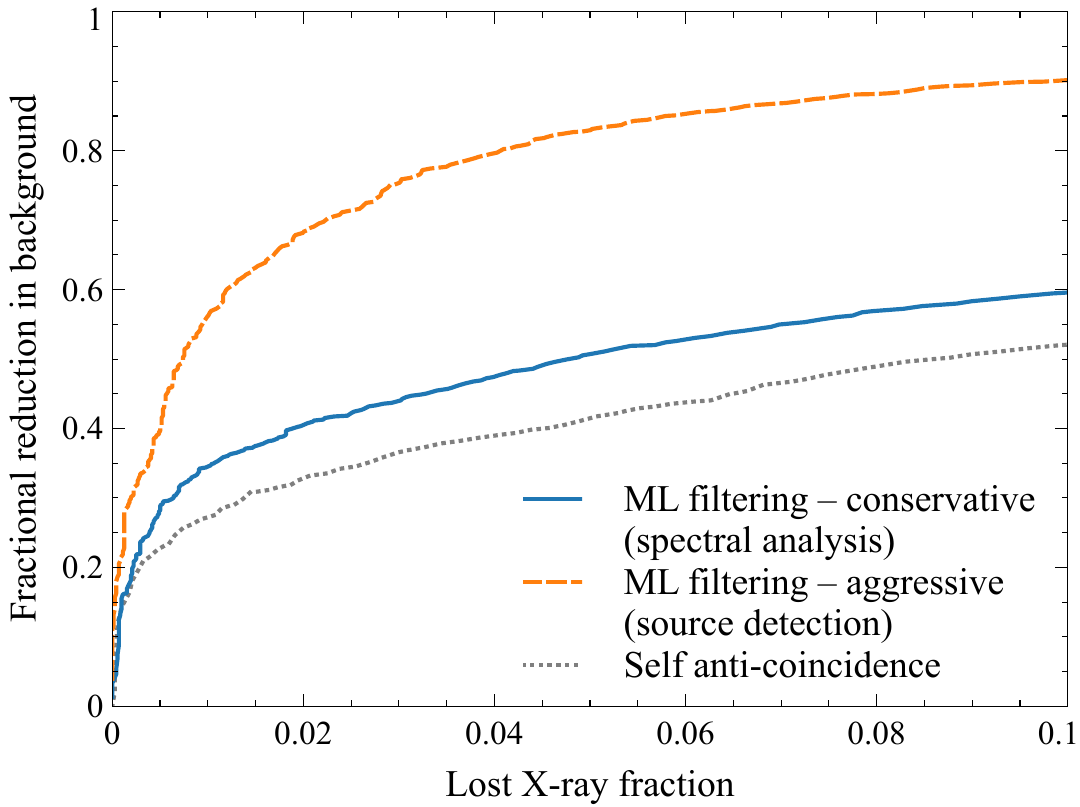}
    }
    \subfigure[] {
    \label{fig:ai_performance:spec}
    \includegraphics[width=0.45\linewidth]{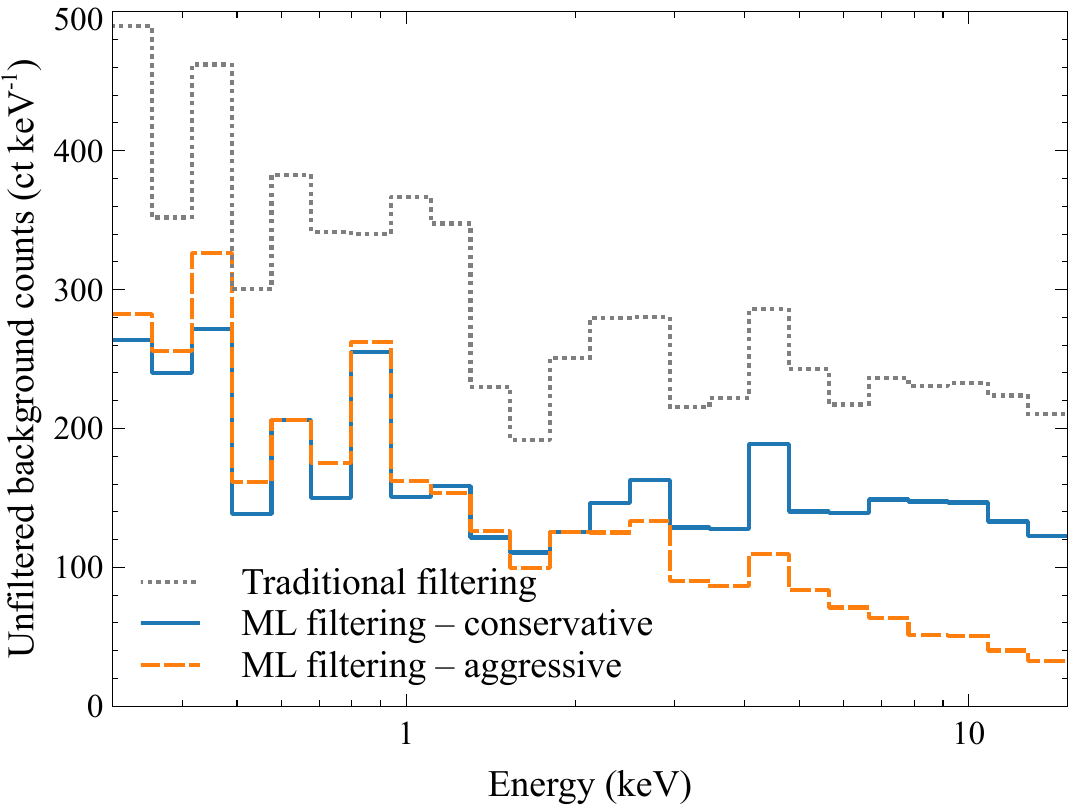}
    }
    \caption{Performance of the AI/ML background filtering algorithm on simulated X-ray and cosmic ray events in a \textit{NewAthena WFI}-like DEPFET detector. \subref{fig:ai_performance:roc} The fractional reduction in background count rate relative to filtering using the traditional method \textit{vs.} the fraction of genuine astrophysical X-ray events that are incorrectly filtered by the algorithm as the threshold classification probability is tuned. Two variants of the ML algorithm are compared: a conservative version suitable for precision spectral analysis, and an aggressive version that provides enhanced background filtering at the expense of spectral calibration that is suited to the detection of faint X-ray sources in deep imaging surveys. The performance of the ML algorithm is also compared to that of the self anti-coincidence filtering method in which any events in pixels within a defined radius of any particle track or invalid event are filtered. \subref{fig:ai_performance:spec} The remaining background spectrum as a function of event energy when filtered using the traditional filtering method, and the conservative and aggressive variants of the ML algorithm.}
    \label{fig:ai_performance}
\end{figure}

The performance of the AI/ML background filtering algorithm was tested upon events in a further set of simulated detector frames that were not included in the training set. In these tests we set the filtering threshold (\textit{i.e.} the cosmic ray classification value produced by the two-stage ML algorithm above which events are rejected) such that the false positive rate (the fraction of genuine X-ray events that are incorrectly filtered) is limited to 2 per cent. The performance of the ML algorithm is shown in Figure~\ref{fig:ai_performance}.

We find that the conservative version of the ML algorithm, which is not able to base the classification solely upon the energy of individual events, thus preserving the shape of the remaining astrophysical X-ray spectrum, is able to reduce the unfiltered background by 40 per cent relative to traditional filtering methods. The aggressive mode of the algorithm, which can be employed in the detection of faint X-ray sources at the expense of the spectral calibration, reduces the unfiltered background by 68 per cent with respect to traditional filtering methods.

An alternative method to filter `valid' secondary events produced by cosmic ray interactions within the spacecraft and detector is \textit{self anti-coincidence} (SAC)\cite{miller_jatis}. The SAC method masks all pixels within a fixed radius of any pixel within either an extended particle track or other invalid background event. SAC is tuned by adjusting the radius of masked region. We find that the ML filtering algorithm provides improved filtering over SAC. Limiting the lost X-ray signal to 2 per cent, SAC provides a 32 per cent reduction in the background relative to traditional filtering methods, compared to the 40 or 68 per cent achieved by conservative and aggressive variants of the ML algorithm. SAC relies on the primary particle track or other high energy charged particle event appearing within the frame, and relies upon the `valid' secondary events appearing within the mask radius. On the other hand, the ML algorithm is able to draw upon the correlations between particle-induced events across the frame, and where multiple `valid' secondaries appear in close proximity to one another, the ML algorithm can in principle learn the event-to-event correlations that identify background events without the need for the high energy primary event appearing within the frame.

\section{NEXT GENERATION X-RAY EVENT RECONSTRUCTION}
\begin{figure}
    \centering
    \includegraphics[width=0.35\linewidth]{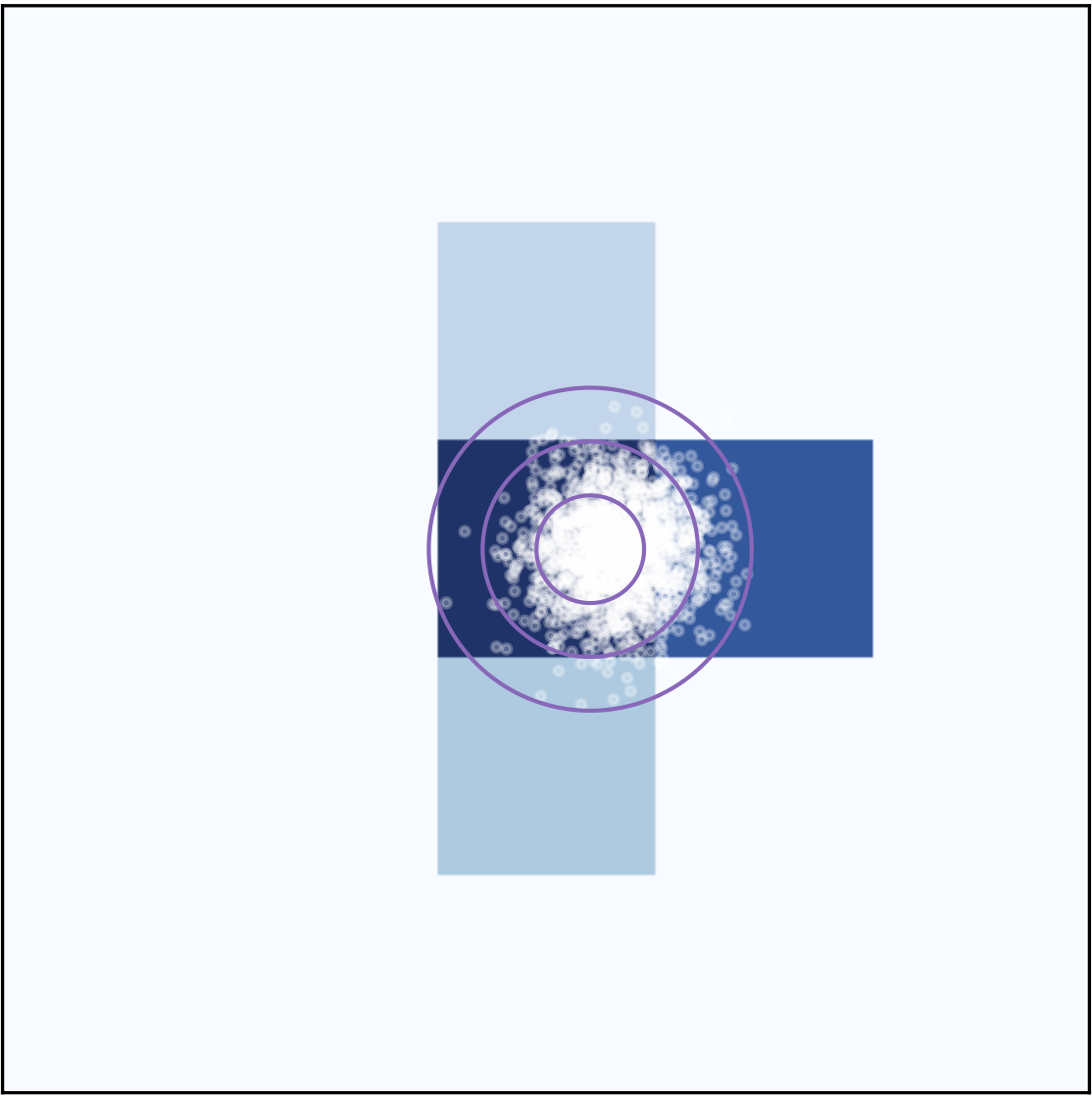}
    \caption{Illustration of the Gaussian fitting event reconstruction algorithm. In this simulated event, the final locations of the electrons following charge diffusion are shown by the white dots. In traditional event reconstruction algorithms, the pixel values indicated by the blue shading are summed. In this next-generation algorithm, a Gaussian model of the charge cloud, shown by the purple contours, is fit to the measured pixel values.}
    \label{fig:event_fit}
\end{figure}

To improve the sensitivity and spectral resolution of pixelated silicon X-ray detectors (including CCDs, CMOS and DEPFET sensors) at energies below 1\,keV, we are developing a new \textit{Gaussian fitting} algorithm to reconstruct X-ray events and combining the signal from neighboring pixels to calculate the energy of split events. Rather than summing the pixels in which the recorded signal exceeds the split threshold\cite{trad_event}, this new algorithm instead fits a physically-motivated model of the charge cloud to the value recorded in each pixel to account for the entire event energy including that lost into those pixels that do not reach the required threshold. Details of the algorithm are presented in Ref.~\citenum{spie_2024} (see also Refs.~\citenum{lamarr+2022,jones+2024}) and the principle of the algorithm is illustrated in Figure~\ref{fig:event_fit}.

The charge cloud itself is modeled as a two-dimensional Gaussian with four free parameters: the amplitude (representing the total number of electrons in the cloud, $N_\mathrm{e}$, and hence the energy of the incoming photon, with one electron produced per 3.65\,eV of incident energy in silicon), the width of the cloud (represented by the standard deviation of the Gaussian profile, $\sigma$), and the location of the event centroid, $(x_0, y_0)$. The value of the model in each pixel is the \textit{integral} of the Gaussian over the pixel area. Critically, fitting the integrated profile over each pixel means that the algorithm is robust to any pixel size, and to the cloud centroid lying at any location relative to the pixel boundaries. For each event, the model is fit to a small island of pixels centered upon the pixel in which the largest pixel value is recorded. In addition to the Gaussian cloud profile, a constant noise term is added to all pixels in the island to account for read noise and other sources of noise inherent in the pixel values. The mean noise level in pixels without photon signal is a further free parameter, $\mu_\mathrm{noise}$, that is fit to the data. The full model expression for the signal in each pixel, $m_{ij}$, with pixel boundaries at $(x_{ij}, y_{ij})$, is:
\begin{equation}
    m_{ij} =  N_\mathrm{e} \left[\mathrm{erf}\left(\frac{x_{i+1} - x_0}{\sqrt{2}\sigma}\right) - \mathrm{erf}\left(\frac{x_{i} - x_0}{\sqrt{2}\sigma}\right)\right]\left[\mathrm{erf}\left(\frac{y_{j+1} - y_0}{\sqrt{2}\sigma}\right) - \mathrm{erf}\left(\frac{y_{j} - y_0}{\sqrt{2}\sigma}\right)\right] + \mu_\mathrm{noise}
\end{equation}

This model is fit to the measured value in each of the pixels within the event island via weighted least squares minimization, in which the fit statistic comparing the model values $m_{ij}$ to the signal in each pixel $s_{ij}$ with RMS noise $\sigma_\mathrm{noise}$ in each measurement, is given by $\sum_{ij} (s_{ij} - m_{ij})/\sigma_{noise}$.

In principle, event reconstruction via fitting models of the charge cloud within the detector can yield further information about the nature of the event, for example distinguishing single photon events from pile-up events in which two or more photons absorbed during a single frame integration in either the same pixel or neighboring pixels are recorded as a single event with their energies combined. Moreover, if the width of the charge cloud can be determined in addition to the event energy from the best-fitting model parameters, further information is available to discriminate background events from genuine astrophysical X-rays, since the size of the charge cloud recorded at the collection gates following diffusion depends upon the absorption depth of the photon in the silicon, the distribution of which is, in-principle, well understood from the variation of the attenuation length as a function of photon energy.\cite{spie_2024}

\subsection{Laboratory validation of the Gaussian fitting event reconstruction algorithm}
The XOC X-ray beamline at Stanford\cite{pan+2025} allows these next-generation event reconstruction algorithms to be tested upon real CCD detectors. The X-ray fluorescence (XRF) source produces specific emission lines by illuminating a target wheel, with an array materials to produce lines at specific energies. We validated the Gaussian fitting algorithm, re-constructing events detected by an MIT-LL CCID-93 p-JFET CCD detector\cite{stueber+2025,bautz+2021,prigozhin+2020} focusing on the K$\alpha$ fluorescent lines of magnesium at 1.25\,keV and oxygen at 0.52\,keV (using a BeO target). The CCID-93 is a front-illuminated (FI) device $512\times 512$ pixels, each of which is 8\,$\mu\mathrm{m}$ in size with an expected depletion depth of 50\,$\mu\mathrm{m}$. The detector operates with RMS read noise of $2.5\,\mathrm{e}^{-}$. In these tests, the event threshold (required in the central pixel of each event) is defined to be 10 times the read noise level and the split threshold is defined to be 3 times the read noise.

\begin{figure}
    \centering
    \subfigure[Mg K$\alpha$ (1.25\,keV)] {
    \label{fig:reconstructed_spec:mg}
    \includegraphics[width=0.45\linewidth]{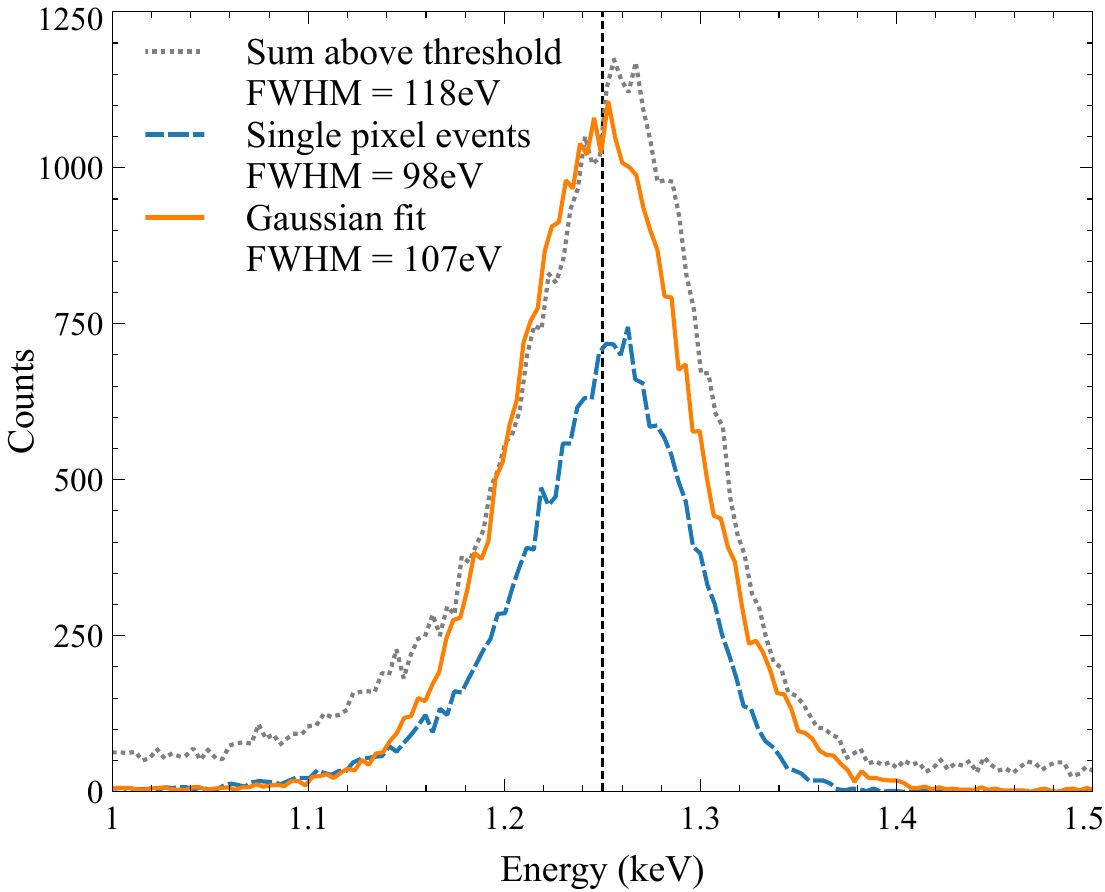}
    }
    \subfigure[O K$\alpha$ (0.52\,keV)] {
    \label{fig:reconstructed_spec:o}
    \includegraphics[width=0.45\linewidth]{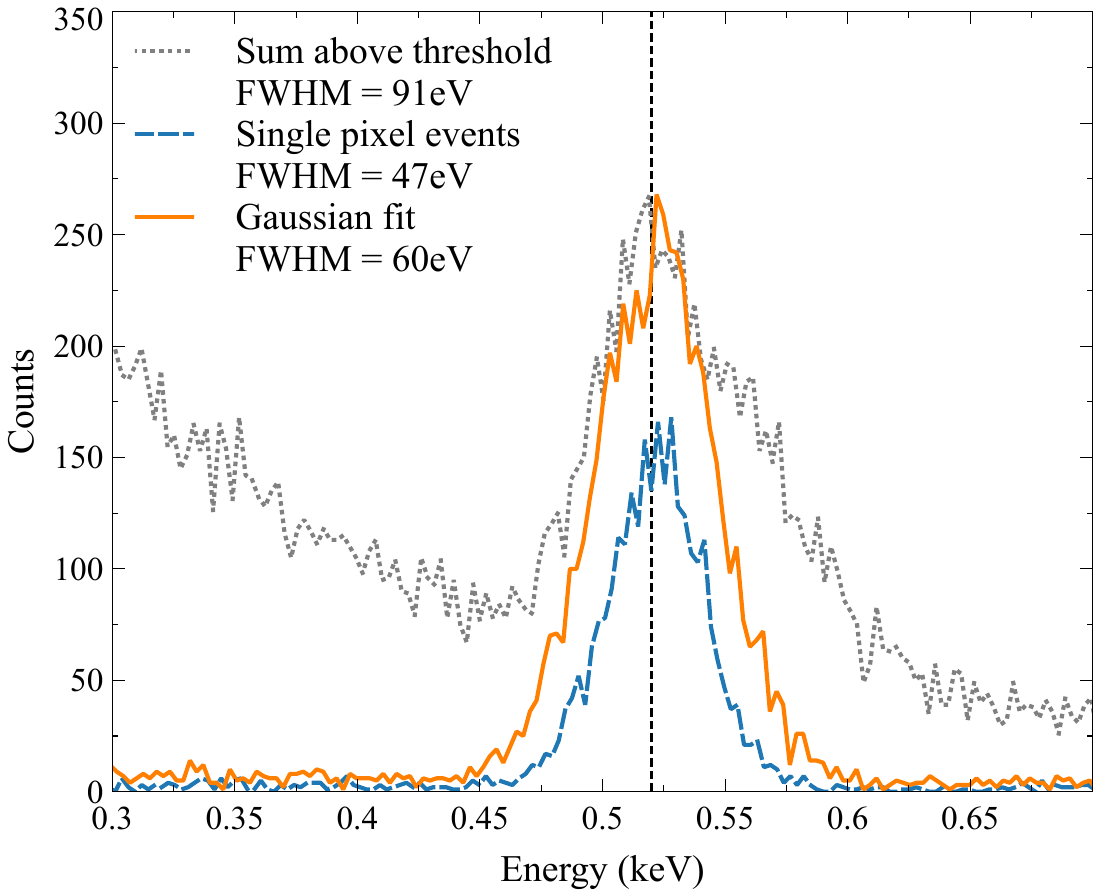}
    }
    \caption{Spectra measured by an MIT-LL CCID-93 CCD detector illuminated by the X-ray fluorescence source in the XOC Beamline at Stanford with \subref{fig:reconstructed_spec:mg} a Mg target, producing the Mg K$\alpha$ fluorescence line at 1.25\,keV, and \subref{fig:reconstructed_spec:o} a BeO target, producing the O K$\alpha$ line at 0.52\,keV. In each case, three event reconstruction techniques are compared: (1) the traditional technique, summing the values recorded in pixels above the split threshold, (2) taking only single pixel events, and (3) the Gaussian fitting technique, fitting a physically-motivated model of the charge cloud produced by the photon in the detector to the recorded pixel values. In each case, the gain (\textit{i.e.} the conversion between ADU detector units and event energy) is calibrated such that that line centroid lies at the true line energy, thus the difference in performance between the three methods can be seen in FWHM of the line spread function obtained in each case.}
    \label{fig:reconstructed_spec}
\end{figure}

The performance of the Gaussian fitting algorithm is compared to the traditional event reconstruction algorithm in Table~\ref{tab:event_reconstruction} for measurements taken when the detector was illuminated using the Mg and BeO target, to produce the Mg K$\alpha$ and O K$\alpha$ lines, respectively. The Gaussian fitting algorithm is compared to the traditional event reconstruction method, summing the pixels in each event that exceed the split threshold. Due to the small pixel size in the CCID-93, a significant number of high-multiplicity events are observed. Since we do not need to filter particle background events in these tests, we include all events, summing all pixels above the split threshold, rather than just selecting single to quadruple pixel events as would be done in-flight. We compare the measured FWHM of the emission line as a proxy for the spectral resolution and the line spread function, in addition to the number of counts detected under the line profile (fit as a Gaussian to the measured spectrum) as a measure of the sensitivity. We also compare to the case when only single pixel (also known as `grade 0' events) are included in the spectrum. The reconstructed spectra using each method are shown in Figure~\ref{fig:reconstructed_spec}.

In each case, the detector gain is calibrated such that the line centroid matches the true energy of the emission line, thus no shift in gain or energy scale conversion is seen between the spectra produced using the different reconstruction methods. Since the CCID-93 is a front-illuminated device, lower energy photons are likely to be absorbed closer to the charge collection gates, thus it is the charge clouds from higher energy events that diffuse to the largest degree. In this case, the degradation of the energy resolution at low photon energies is simply due to the counting statistics as clouds with smaller numbers of electrons are split where events hit closer to the pixel boundaries.

\begin{table}[ht]
\caption{The measured full width at half maximum (FWHM) and integrated counts of the Mg K$\alpha$ and O K$\alpha$ emission lines from an MIT-LL CCID-93 CCD detector, illuminated by the X-ray fluorescence source in the XOC Beamline at Stanford using, respectively a Mg and BeO target. Measurements of the line were obtained when events are reconstructed using the traditional method, summing the pixels above the split threshold, taking only single-pixel events, and reconstructing each event using the Gaussian fitting method, fitting the charge cloud model to islands of different sizes around the brightest event pixel.} 
\label{tab:event_reconstruction}
\begin{center}       
\begin{tabular}{|l|c|c|c|c|}
\hline
\rule[-1ex]{0pt}{3.5ex} & \multicolumn{2}{|c|}{Mg K$\alpha$ (1.25\,keV)} & \multicolumn{2}{|c|}{O K$\alpha$ (0.52\,keV)}  \\
\hline
\rule[-1ex]{0pt}{3.5ex}  & FWHM (eV) & Line counts & FWHM (eV) & Line counts \\
\hline
\rule[-1ex]{0pt}{3.5ex} Sum pixels above threshold  & $117.6 \pm 1.9$ & 36,508 & $91\pm 6$ & 12,992\\
\rule[-1ex]{0pt}{3.5ex} Single pixel events only & $98.3 \pm 0.8$ & 19,403 & $46.7\pm0.6$ & 2,741\\
\hline
\rule[-1ex]{0pt}{3.5ex} Gaussian fit $3\times 3$ & $122.3\pm 0.8$ & 34,454 & $71.8\pm 1.2$ & 6,761\\
\rule[-1ex]{0pt}{3.5ex} Gaussian fit $5\times 5$ & $108.9\pm 0.4$ & 34,019 & $61.7\pm 0.7$ & 6,144\\
\rule[-1ex]{0pt}{3.5ex} Gaussian fit $7\times 7$ & $107.1\pm 0.6$ & 33,588 & $59.7\pm 0.7$ & 5,911\\
\rule[-1ex]{0pt}{3.5ex} Gaussian fit $9\times 9$ & $106.8\pm 0.4$ & 33,080 & $59.9\pm 0.7$ & 5,647\\
\hline
\end{tabular}
\end{center}
\end{table} 

In reconstructing events using the Gaussian fitting algorithm, we trialed fitting the model to different pixel island sizes surrounding the brightest pixel. Trialing $3\times 3$, $5\times 5$, $7\times 7$ and $9\times 9$ pixel islands, we find that fitting events using a larger island improves the spectral resolution (decreasing the FWHM of the lines in the reconstructed spectrum). The larger window allows a more precise constraint to be placed on the noise parameter by fitting pixels far from the event centroid. The improvement is most notable increasing the island size from $3\times 3$ to $5\times 5$, but improves marginally as the island size is increased further. For higher energy photon events (\textit{i.e.} the 1.25\,keV Mg K$\alpha$ line), the $3\times 3$ window is insufficient to accurately determine the noise level, and indeed provides worse energy resolution than the traditional method.

We see that the Gaussian fitting event reconstruction algorithm significantly improves the energy resolution at the lowest photon energies. At 1.25\,keV, the FWHM of the spectral line spread function improves from 117.6\,eV to 106.8\,eV (fitting the charge cloud model to $9\times 9$ pixel islands). At 0.52\,keV, the improvement is more significant as a greater fraction of the energy from lower energy photons is lost into pixels that do not reach the split threshold. The FWHM of the O K$\alpha$ line improves from 91\,eV to 59.9\,eV using the Gaussian fitting method, which is able to account for this `lost' photon energy by fitting the charge cloud model to the measured pixel values.

The best energy resolution is always found when reconstructing the spectrum only from single pixel events, where the photon is absorbed sufficiently far from the pixel boundary that little of the event energy is lost into neighboring pixels. Taking only these events, however, reduces the quantum efficiency of the detector by a factor of between 2 and 5. The Gaussian reconstruction method recovers the spectral resolution but while retaining much of the instrument sensitivity (92 per cent of the line counts are maintained at 1.25\,keV, and 43 per cent of the line counts are maintained at 0.52\,keV, compared to just 20 per cent when taking only single pixel events).  

\section{CONCLUSIONS}
Augmenting next-generation X-ray imaging detectors with AI/ML and advanced event reconstruction algorithms enhances the sensitivity of these detectors to low surface brightness and soft or high-redshift X-ray sources, addressing two key limitations.

Machine learning algorithms provide enhanced filtering of the cosmic ray-induced instrumental background by considering every event in the context within which it appears in frames read out from the detector. We have developed a prototype two-stage ML algorithm, combining a convolutional neural network (CNN) with a random forest classifier that reduces the unfiltered background by 40 per cent with respect to traditional filtering methods in a conservative mode that preserves the spectral calibration of the final event data. An aggressive mode of the filtering algorithm was also trained, incorporating additional information about the energy of isolated events. This aggressive variant provides a further enhancement, reducing the unfiltered background by 68 per cent relative to traditional methods, but at the expense of precise spectral calibration.

Advanced event reconstruction algorithms enhance the sensitivity and energy resolution of pixelated silicon X-ray imaging detectors by fitting a physically-motivated model of the charge cloud generated by photons within the detector to the values recorded in pixels. We have demonstrated the performance of such an algorithm on an MIT-LL CCID-93 CCD detector in a laboratory setting. This Gaussian fitting algorithm significantly improves the energy resolution at low photon energies, decreasing the FWHM of the line spread function of the O K$\alpha$ line at 0.52\,keV from 91\,eV (using the traditional reconstruction method) to 60\,eV. We find that event energy determination is improved when a small ($9\times 9$) island of empty pixels is included around each event, placing a tighter constraint on the noise term in the charge cloud model.

While current prototypes of these algorithms are designed to run in the ground segment, future developments will focus on compressing these algorithms into a form that can be run upon lightweight computing hardware on board the spacecraft, alongside the detector, receiving the raw pixel-by-pixel data stream, in an `edge-AI' setting. Further development of the Gaussian fitting event reconstruction algorithm will provide additional event information to allow for enhanced identification of pile-up and particle-induced background events, and characterization of the distortion of charge clouds that arises from processes beyond diffusion, including charge transfer inefficiency (CTI). These algorithms can be developed in tandem with the detectors themselves, to identify the optimal pixel architecture (including pitch and depth) for the measurement of the charge cloud properties. Combined with the capabilities of next-generation high-speed, low-noise detectors, these algorithms pave the way to satisfying the requirements for future X-ray flagship missions.

\acknowledgments 
This work has been supported by the NASA \textit{Astrophysics Research and Analysis} (APRA) program under grant number 80NSSC22K0342. 

\bibliography{aibkg} 

@INPROCEEDINGS{heine+2026,
       author = {{Heine}, Matthew K. and {Grant}, Catherine E. and {Bautz}, Marshall W. and {LaMarr}, Beverly J. and {Miller}, Eric D. and {Hubbard}, Michael W.J. and {Hall}, David and {Requena}, Joan and {Perinati}, Emanuele and {Allen}, Steven W. and {Poliszczuk}, Artem and {Wilkins}, Dan and {Gastaldello}, Fabio and {Molendi}, Silvano and {Kraft}, Ralph P. and {Schellenberger}, Gerrit and {Sarkar}, Arnab},
        title = "{High-statistics simulations of NewAthena WFI background using Geant4}",
    booktitle = {Space Telescopes and Instrumentation 2026: Ultraviolet to Gamma Ray},
         year = 2026,
       editor = {{Nikzad}, Shouleh and {Nakazawa}, Kazuhiro and {Feroci}, Marco},
       series = {Society of Photo-Optical Instrumentation Engineers (SPIE) Conference Series},
       volume = {14146},
          eid = {14146177},
        pages = {14146177},
          doi = {},
 archivePrefix = {arXiv},
       eprint = {2607.26163},
 primaryClass = {astro-ph.IM},
      }

@ARTICLE{trad_event,
       author = {{Lumb}, David H. and {Holland}, Andrew D.},
        title = "{Event recognition techniques in CCD X-ray detectors for astronomy}",
      journal = {Nuclear Instruments and Methods in Physics Research A},
         year = 1988,
        month = dec,
       volume = {273},
       number = {2-3},
        pages = {696-700},
          doi = {10.1016/0168-9002(88)90081-2},
       adsurl = {https://ui.adsabs.harvard.edu/abs/1988NIMPA.273..696L}
}

@article{random_forest,
	title = {Random {Forests}},
	volume = {45},
	issn = {1573-0565},
	url = {https://doi.org/10.1023/A:1010933404324},
	doi = {10.1023/A:1010933404324},
	number = {1},
	journal = {Machine Learning},
	author = {Breiman, Leo},
	month = oct,
	year = {2001},
	pages = {5--32},
}

@INPROCEEDINGS{resnet,
  author={He, Kaiming and Zhang, Xiangyu and Ren, Shaoqing and Sun, Jian},
  booktitle={2016 IEEE Conference on Computer Vision and Pattern Recognition (CVPR)}, 
  title={Deep Residual Learning for Image Recognition}, 
  year={2016},
  volume={},
  number={},
  pages={770-778},
  doi={10.1109/CVPR.2016.90}}

@INPROCEEDINGS {cam,
author = {B. Zhou and A. Khosla and A. Lapedriza and A. Oliva and A. Torralba},
booktitle = {2016 IEEE Conference on Computer Vision and Pattern Recognition (CVPR)},
title = {Learning Deep Features for Discriminative Localization},
year = {2016},
volume = {},
issn = {1063-6919},
pages = {2921-2929},
doi = {10.1109/CVPR.2016.319},
url = {https://doi.ieeecomputersociety.org/10.1109/CVPR.2016.319},
publisher = {IEEE Computer Society},
address = {Los Alamitos, CA, USA},
month = {jun}
}

@ARTICLE{sisero_jatis_2,
       author = {{Chattopadhyay}, Tanmoy and {Herrmann}, Sven and {Orel}, Peter and {Donlon}, Kevan and {Prigozhin}, Gregory and {Morris}, Glenn and {Cooper}, Michael and {LaMarr}, Beverly and {Malonis}, Andrew and {Allen}, Steven W. and {Bautz}, Marshall W. and {Leitz}, Chris},
        title = "{Demonstrating repetitive non-destructive readout with SiSeRO devices}",
      journal = {Journal of Astronomical Telescopes, Instruments, and Systems},
         year = 2024,
        month = jan,
       volume = {10},
          eid = {016004},
        pages = {016004},
          doi = {10.1117/1.JATIS.10.1.016004},
archivePrefix = {arXiv},
       eprint = {2305.01900},
 primaryClass = {astro-ph.IM},
       adsurl = {https://ui.adsabs.harvard.edu/abs/2024JATIS..10a6004C}
}

@ARTICLE{sisero_jatis_1,
       author = {{Chattopadhyay}, Tanmoy and {Herrmann}, Sven and {Burke}, Barry E. and {Donlon}, Kevan and {Prigozhin}, Gregory and {Morris}, Glenn and {Orel}, Peter and {Cooper}, Michael and {Malonis}, Andrew and {Wilkins}, Daniel R. and {Suntharalingam}, Vyshnavi and {Allen}, Steven W. and {Bautz}, Marshall W. and {Leitz}, Chris},
        title = "{First results on SiSeRO devices: a new x-ray detector for scientific instrumentation}",
      journal = {Journal of Astronomical Telescopes, Instruments, and Systems},
         year = 2022,
        month = apr,
       volume = {8},
          eid = {026006},
        pages = {026006},
          doi = {10.1117/1.JATIS.8.2.026006},
archivePrefix = {arXiv},
       eprint = {2112.05033},
 primaryClass = {astro-ph.IM},
       adsurl = {https://ui.adsabs.harvard.edu/abs/2022JATIS...8b6006C}
}

@INPROCEEDINGS{bautz+2021,
       author = {{Bautz}, M. and {Burke}, B. and {Cooper}, M. and {Craig}, D. and {Donlon}, K. and {Foster}, R. and {Grant}, C.~E. and {LaMarr}, B. and {Leitz}, C. and {Malonis}, A. and {Miller}, E. and {Prigozhin}, G. and {Thayer}, C. and {Allen}, S. and {Herrmann}, S. and {Chattopadhyay}, T. and {Morris}, R.~G.},
        title = "{Progress toward fast, low-noise, low-power CCDs for Lynx and other high-energy astrophysics missions}",
    booktitle = {Society of Photo-Optical Instrumentation Engineers (SPIE) Conference Series},
         year = 2021,
       editor = {{den Herder}, Jan-Willem A. and {Nikzad}, Shouleh and {Nakazawa}, Kazuhiro},
       series = {Society of Photo-Optical Instrumentation Engineers (SPIE) Conference Series},
       volume = {11444},
        month = jan,
          eid = {1144494},
        pages = {1144494},
          doi = {10.1117/12.2561441},
       adsurl = {https://ui.adsabs.harvard.edu/abs/2021SPIE11444E..94B}
}

@INPROCEEDINGS{prigozhin+2020,
       author = {{Prigozhin}, G. and {Burke}, B. and {Cooper}, M. and {Donlon}, K. and {Leitz}, C. and {Foster}, R. and {LaMarr}, B. and {Malonis}, A. and {Bautz}, M.},
        title = "{Charge transfer effects in a CCD with a single polysilicon gate structure}",
    booktitle = {X-Ray, Optical, and Infrared Detectors for Astronomy IX},
         year = 2020,
       editor = {{Holland}, Andrew D. and {Beletic}, James},
       series = {Society of Photo-Optical Instrumentation Engineers (SPIE) Conference Series},
       volume = {11454},
        month = dec,
          eid = {114541C},
        pages = {114541C},
          doi = {10.1117/12.2562677},
       adsurl = {https://ui.adsabs.harvard.edu/abs/2020SPIE11454E..1CP}
}

@INPROCEEDINGS{spie_2024,
       author = {{Wilkins}, D.~R. and {Poliszczuk}, A. and {Schneider}, B. and {Miller}, E.~D. and {Allen}, S.~W. and {Bautz}, M. and {Chattopadhyay}, T. and {Falcone}, A.~D. and {Foster}, R. and {Grant}, C.~E. and {Herrmann}, S. and {Kraft}, R. and {Morris}, R.~G. and {Nulsen}, P. and {Orel}, P. and {Schellenberger}, G.},
        title = "{Augmenting astronomical x-ray detectors with AI for enhanced sensitivity and reduced background}",
    booktitle = {Space Telescopes and Instrumentation 2024: Ultraviolet to Gamma Ray},
         year = 2024,
       editor = {{den Herder}, Jan-Willem A. and {Nikzad}, Shouleh and {Nakazawa}, Kazuhiro},
       series = {Society of Photo-Optical Instrumentation Engineers (SPIE) Conference Series},
       volume = {13093},
        month = aug,
          eid = {130931R},
        pages = {130931R},
          doi = {10.1117/12.3019396},
       adsurl = {https://ui.adsabs.harvard.edu/abs/2024SPIE13093E..1RW}
}

@INPROCEEDINGS{pan+2025,
       author = {{Pan}, Abigail Y. and {Stueber}, Haley R. and {Chattopadhyay}, Tanmoy and {Allen}, Steven W. and {Bautz}, Marshall W. and {Donlon}, Kevan and {Grant}, Catherine E. and {Herrmann}, Sven and {LaMarr}, Beverly and {Malonis}, Andrew and {Miller}, Eric D. and {Morris}, Glenn and {Orel}, Peter and {Poliszczuk}, Artem and {Prigozhin}, Gregory and {Wilkins}, Dan},
        title = "{Design, development, and commissioning of a flexible test setup for the AXIS prototype detector}",
    booktitle = {UV, X-Ray, and Gamma-Ray Space Instrumentation for Astronomy XXIV},
         year = 2025,
       editor = {{Siegmund}, Oswald H. and {Hoadley}, Keri},
       series = {Society of Photo-Optical Instrumentation Engineers (SPIE) Conference Series},
       volume = {13625},
        month = sep,
          eid = {136251J},
        pages = {136251J},
          doi = {10.1117/12.3063468},
archivePrefix = {arXiv},
       eprint = {2508.14177},
 primaryClass = {astro-ph.IM},
       adsurl = {https://ui.adsabs.harvard.edu/abs/2025SPIE13625E..1JP}
}

@INPROCEEDINGS{stueber+2025,
       author = {{Stueber}, Haley R. and {Pan}, Abigail Y. and {Chattopadhyay}, Tanmoy and {Allen}, Steven W. and {Bautz}, Marshall W. and {Donlon}, Kevan and {Grant}, Catherine E. and {Herrmann}, Sven and {LaMarr}, Beverly J. and {Malonis}, Andrew and {Miller}, Eric D. and {Morris}, R. Glenn and {Orel}, Peter and {Poliszczuk}, Artem and {Prigozhin}, Gregory Y. and {Wilkins}, Daniel R.},
        title = "{Fast, low noise CCD systems for future strategic x-ray missions}",
    booktitle = {UV, X-Ray, and Gamma-Ray Space Instrumentation for Astronomy XXIV},
         year = 2025,
       editor = {{Siegmund}, Oswald H. and {Hoadley}, Keri},
       series = {Society of Photo-Optical Instrumentation Engineers (SPIE) Conference Series},
       volume = {13625},
        month = sep,
          eid = {136251Q},
        pages = {136251Q},
          doi = {10.1117/12.3064612},
archivePrefix = {arXiv},
       eprint = {2508.14174},
 primaryClass = {astro-ph.IM},
       adsurl = {https://ui.adsabs.harvard.edu/abs/2025SPIE13625E..1QS}
}

@INPROCEEDINGS{poliszczuk+2024,
       author = {{Poliszczuk}, Artem and {Wilkins}, Dan and {Allen}, Steven W. and {Miller}, Eric D. and {Chattopadhyay}, Tanmoy and {Schneider}, Benjamin and {Darve}, Julien Eric and {Bautz}, Marshall and {Falcone}, Abe and {Foster}, Richard and {Grant}, Catherine E. and {Herrmann}, Sven and {Kraft}, Ralph and {Morris}, R. Glenn and {Nulsen}, Paul and {Orel}, Peter and {Schellenberger}, Gerrit and {Stueber}, Haley R.},
        title = "{Towards efficient machine-learning-based reduction of the cosmic-ray induced background in x-ray imaging detectors: increasing context awareness}",
    booktitle = {Space Telescopes and Instrumentation 2024: Ultraviolet to Gamma Ray},
         year = 2024,
       editor = {{den Herder}, Jan-Willem A. and {Nikzad}, Shouleh and {Nakazawa}, Kazuhiro},
       series = {Society of Photo-Optical Instrumentation Engineers (SPIE) Conference Series},
       volume = {13093},
        month = aug,
          eid = {130931T},
        pages = {130931T},
          doi = {10.1117/12.3020598},
       adsurl = {https://ui.adsabs.harvard.edu/abs/2024SPIE13093E..1TP}
}

@article{xis,
    author = {Koyama, Katsuji and Tsunemi, Hiroshi and Dotani, Tadayasu and Bautz, Mark W. and Hayashida, Kiyoshi and Tsuru, Takeshi Go and Matsumoto, Hironori and Ogawara, Yoshiaki and Ricker, George R. and Doty, John and Kissel, Steven E. and Foster, Rick and Nakajima, Hiroshi and Yamaguchi, Hiroya and Mori, Hideyuki and Sakano, Masaaki and Hamaguchi, Kenji and Nishiuchi, Mamiko and Miyata, Emi and Torii, Ken’ichi and Namiki, Masaaki and Katsuda, Satoru and Matsuura, Daisuke and Miyauchi, Tomofumi and Anabuki, Naohisa and Tawa, Noriaki and Ozaki, Masanobu and Murakami, Hiroshi and Maeda, Yoshitomo and Ichikawa, Yoshinori and Prigozhin, Gregory Y. and Boughan, Edward A. and LaMarr, Beverly and Miller, Eric D. and Burke, Barry E. and Gregory, James A. and Pillsbury, Allen and Bamba, Aya and Hiraga, Junko S. and Senda, Atsushi and Katayama, Haruyoshi and Kitamoto, Shunji and Tsujimoto, Masahiro and Kohmura, Takayoshi and Tsuboi, Yohko and Awaki, Hisamitsu},
    title = "{X-Ray Imaging Spectrometer (XIS) on Board Suzaku}",
    journal = {Publications of the Astronomical Society of Japan},
    volume = {59},
    number = {sp1},
    pages = {S23-S33},
    year = {2007},
    month = {01},
    issn = {0004-6264},
    doi = {10.1093/pasj/59.sp1.S23},
    url = {https://doi.org/10.1093/pasj/59.sp1.S23},
    eprint = {https://academic.oup.com/pasj/article-pdf/59/sp1/S23/17451643/pasj59-0S23.pdf},
}

@inproceedings{poliszczuk+2023,
	author = {Artem Poliszczuk and Dan Wilkins and Steven W. Allen and Eric Miller and Tanmoy Chattopadhyay and Marshall Bautz and Julien Eric Darve and Richard Foster and Catherine E. Grant and Sven Herrmann and Ralph Kraft and R. Glenn Morris and Peter Orel and Arnab Sarkar and Benjamin Schneider},
	booktitle = {Applications of Machine Learning 2023},
	doi = {10.1117/12.2677095},
	editor = {Michael E. Zelinski and Tarek M. Taha and Jonathan Howe and Barath Narayanan Narayanan},
	organization = {International Society for Optics and Photonics},
	pages = {126750C},
	publisher = {SPIE},
	title = {{Reduction of cosmic-ray induced background in astronomical x-ray imaging detectors via image segmentation methods}},
	url = {https://doi.org/10.1117/12.2677095},
	volume = {12675},
	year = {2023}}

@ARTICLE{miller_jatis,
       author = {{Miller}, Eric D. and {Grant}, Catherine E. and {Bautz}, Marshall W. and {Molendi}, Silvano and {Kraft}, Ralph and {Nulsen}, Paul and {Bulbul}, Esra and {Allen}, Steven and {Burrows}, David N. and {Eraerds}, Tanja and {Fioretti}, Valentina and {Gastaldello}, Fabio and {Hall}, David and {Hubbard}, Michael W.~J. and {Keelan}, Jonathan and {Meidinger}, Norbert and {Perinati}, Emanuele and {Rau}, Arne and {Wilkins}, Dan},
        title = "{Mitigating the effects of particle background on the Athena Wide Field Imager}",
      journal = {Journal of Astronomical Telescopes, Instruments, and Systems},
         year = 2022,
        month = jan,
       volume = {8},
          eid = {018001},
        pages = {018001},
          doi = {10.1117/1.JATIS.8.1.018001},
archivePrefix = {arXiv},
       eprint = {2202.00064},
 primaryClass = {astro-ph.IM},
       adsurl = {https://ui.adsabs.harvard.edu/abs/2022JATIS...8a8001M}
}

@ARTICLE{poissonccd,
       author = {{Lage}, Craig and {Bradshaw}, Andrew and {Anthony Tyson}, J. and {LSST Dark Energy Science Collaboration}},
        title = "{Poisson\_CCD: A dedicated simulator for modeling CCDs}",
      journal = {Journal of Applied Physics},
         year = 2021,
        month = oct,
       volume = {130},
       number = {16},
          eid = {164502},
        pages = {164502},
          doi = {10.1063/5.0058894},
archivePrefix = {arXiv},
       eprint = {1911.09038},
 primaryClass = {astro-ph.IM},
       adsurl = {https://ui.adsabs.harvard.edu/abs/2021JAP...130p4502L}
}

@INPROCEEDINGS{miller+2022,
       author = {{Miller}, Eric D. and {Prigozhin}, Gregory Y. and {LaMarr}, Beverly J. and {Bautz}, Marshall W. and {Foster}, Richard F. and {Grant}, Catherine E. and {Lage}, Craig S. and {Leitz}, Christopher and {Malonis}, Andrew},
        title = "{Understanding the effects of charge diffusion in next-generation soft x-ray imagers}",
    booktitle = {Space Telescopes and Instrumentation 2022: Ultraviolet to Gamma Ray},
         year = 2022,
       editor = {{den Herder}, Jan-Willem A. and {Nikzad}, Shouleh and {Nakazawa}, Kazuhiro},
       series = {Society of Photo-Optical Instrumentation Engineers (SPIE) Conference Series},
       volume = {12181},
        month = aug,
          eid = {121816R},
        pages = {121816R},
          doi = {10.1117/12.2628929},
archivePrefix = {arXiv},
       eprint = {2208.07348},
 primaryClass = {astro-ph.IM},
       adsurl = {https://ui.adsabs.harvard.edu/abs/2022SPIE12181E..6RM}
}

@ARTICLE{jones+2024,
       author = {{Jones}, L.~S. and {Crews}, C. and {Ivory}, J. and {Holland}, A.},
        title = "{A model for photon counting X-ray event reconstruction uncertainty}",
      journal = {Journal of Instrumentation},
         year = 2024,
        month = may,
       volume = {19},
       number = {5},
          eid = {P05017},
        pages = {P05017},
          doi = {10.1088/1748-0221/19/05/P05017},
       adsurl = {https://ui.adsabs.harvard.edu/abs/2024JInst..19P5017J}
}

@ARTICLE{lamarr+2022,
       author = {{LaMarr}, Beverly J. and {Prigozhin}, Gregory Y. and {Miller}, Eric D. and {Thayer}, Carolyn and {Bautz}, Marshall W. and {Foster}, Richard and {Grant}, Catherine E. and {Malonis}, Andrew and {Burke}, Barry E. and {Cooper}, Michael and {Donlon}, Kevan and {Leitz}, Christopher},
        title = "{Measurement and simulation of charge diffusion in a small-pixel charge-coupled device}",
      journal = {Journal of Astronomical Telescopes, Instruments, and Systems},
         year = 2022,
        month = jan,
       volume = {8},
          eid = {016004},
        pages = {016004},
          doi = {10.1117/1.JATIS.8.1.016004},
archivePrefix = {arXiv},
       eprint = {2201.07645},
 primaryClass = {astro-ph.IM},
       adsurl = {https://ui.adsabs.harvard.edu/abs/2022JATIS...8a6004L}
}

@INPROCEEDINGS{xray_speed,
       author = {{Herrmann}, S. and {Orel}, P. and {Chattopadhyay}, T. and {Morris}, R.~G. and {Prigozhin}, G. and {Donlon}, K. and {Foster}, R. and {Bautz}, M. and {Allen}, S. and {Leitz}, C.},
        title = "{X-ray speed reading: enabling fast, low noise readout for next-generation CCDs}",
    booktitle = {X-Ray, Optical, and Infrared Detectors for Astronomy X},
         year = 2022,
       editor = {{Holland}, Andrew D. and {Beletic}, James},
       series = {Society of Photo-Optical Instrumentation Engineers (SPIE) Conference Series},
       volume = {12191},
        month = aug,
          eid = {1219117},
        pages = {1219117},
          doi = {10.1117/12.2630195},
archivePrefix = {arXiv},
       eprint = {2208.01247},
 primaryClass = {astro-ph.IM},
       adsurl = {https://ui.adsabs.harvard.edu/abs/2022SPIE12191E..17H}
}

@INPROCEEDINGS{spie_2022,
       author = {{Wilkins}, D.~R. and {Allen}, S.~W. and {Miller}, E.~D. and {Bautz}, M. and {Chattopadhyay}, T. and {Foster}, R. and {Grant}, C.~E. and {Herrmann}, S. and {Kraft}, R. and {Morris}, R.~G. and {Nulsen}, P. and {Schellenberger}, G.},
        title = "{Reducing the background in x-ray imaging detectors via machine learning}",
    booktitle = {Space Telescopes and Instrumentation 2022: Ultraviolet to Gamma Ray},
         year = 2022,
       editor = {{den Herder}, Jan-Willem A. and {Nikzad}, Shouleh and {Nakazawa}, Kazuhiro},
       series = {Society of Photo-Optical Instrumentation Engineers (SPIE) Conference Series},
       volume = {12181},
        month = aug,
          eid = {121816S},
        pages = {121816S},
          doi = {10.1117/12.2629496},
archivePrefix = {arXiv},
       eprint = {2208.07906},
 primaryClass = {astro-ph.IM},
       adsurl = {https://ui.adsabs.harvard.edu/abs/2022SPIE12181E..6SW}
}

@INPROCEEDINGS{spie_2020,
       author = {{Wilkins}, D.~R. and {Allen}, S.~W. and {Miller}, E.~D. and {Bautz}, M. and {Chattopadhyay}, T. and {Fort}, S. and {Grant}, C.~E. and {Herrmann}, S. and {Kraft}, R. and {Morris}, R.~G. and {Nulsen}, P.},
        title = "{Identifying charged particle background events in x-ray imaging detectors with novel machine learning algorithms}",
    booktitle = {Society of Photo-Optical Instrumentation Engineers (SPIE) Conference Series},
         year = 2020,
       series = {Society of Photo-Optical Instrumentation Engineers (SPIE) Conference Series},
       volume = {11444},
        month = dec,
          eid = {114442O},
        pages = {114442O},
          doi = {10.1117/12.2562354},
archivePrefix = {arXiv},
       eprint = {2012.01463},
 primaryClass = {astro-ph.IM},
       adsurl = {https://ui.adsabs.harvard.edu/abs/2020SPIE11444E..2OW}
}

@INPROCEEDINGS{grant+2020,
       author = {{Grant}, Catherine E. and {Miller}, Eric D. and {Bautz}, Marshall W. and {Eraerds}, Tanja and {Molendi}, Silvano and {Keelan}, Jonathan and {Hall}, David and {Holland}, Andrew D. and {Kraft}, Ralph and {Bulbul}, Esra and {Nulsen}, Paul and {Allen}, Steven},
        title = "{Reducing the Athena WFI charged particle background: results from Geant4 simulations}",
    booktitle = {Society of Photo-Optical Instrumentation Engineers (SPIE) Conference Series},
         year = 2020,
       series = {Society of Photo-Optical Instrumentation Engineers (SPIE) Conference Series},
       volume = {11444},
        month = dec,
          eid = {1144442},
        pages = {1144442},
          doi = {10.1117/12.2561124},
archivePrefix = {arXiv},
       eprint = {2012.01347},
 primaryClass = {astro-ph.IM},
       adsurl = {https://ui.adsabs.harvard.edu/abs/2020SPIE11444E..42G}
}

@INPROCEEDINGS{miller_charge_diff,
       author = {{Miller}, Eric D. and {Foster}, Richard and {Lage}, Craig and {Prigozhin}, Gregory and {Bautz}, Marshall and {Grant}, Catherine and {LaMarr}, Beverly and {Malonis}, Andrew},
        title = "{The effects of charge diffusion on soft x-ray response for future high-resolution imagers}",
    booktitle = {Space Telescopes and Instrumentation 2018: Ultraviolet to Gamma Ray},
         year = 2018,
       editor = {{den Herder}, Jan-Willem A. and {Nikzad}, Shouleh and {Nakazawa}, Kazuhiro},
       series = {Society of Photo-Optical Instrumentation Engineers (SPIE) Conference Series},
       volume = {10699},
        month = jul,
          eid = {106995R},
        pages = {106995R},
          doi = {10.1117/12.2313565},
       adsurl = {https://ui.adsabs.harvard.edu/abs/2018SPIE10699E..5RM}
}

@INPROCEEDINGS{lynx_science,
       author = {{Bautz}, M.~W.},
        title = "{The Lynx X-Ray Observatory: Science Drivers}",
    booktitle = {UV, X-Ray, and Gamma-Ray Space Instrumentation for Astronomy XXI},
         year = 2019,
       series = {Society of Photo-Optical Instrumentation Engineers (SPIE) Conference Series},
       volume = {11118},
        month = sep,
          eid = {111180J},
        pages = {111180J},
          doi = {10.1117/12.2534555},
       adsurl = {https://ui.adsabs.harvard.edu/abs/2019SPIE11118E..0JB}
}

@INPROCEEDINGS{athena_science,
       author = {{Rau}, Arne and {Nandra}, Kirpal and {Aird}, James and {Comastri}, Andrea and {Dauser}, Thomas and {Merloni}, Andrea and {Pratt}, Gabriel W. and {Reiprich}, Thomas H. and {Fabian}, Andy C. and {Georgakakis}, Antonis and {G{\"u}del}, Manuel and {R{\'o}{\.Z}a{\'n}ska}, Agata and {Sanders}, Jeremy S. and {Sasaki}, Manami and {Vaughan}, Simon and {Wilms}, J{\"o}rn and {Meidinger}, Norbert},
        title = "{Athena Wide Field Imager key science drivers}",
    booktitle = {Space Telescopes and Instrumentation 2016: Ultraviolet to Gamma Ray},
         year = 2016,
       editor = {{den Herder}, Jan-Willem A. and {Takahashi}, Tadayuki and {Bautz}, Marshall},
       series = {Society of Photo-Optical Instrumentation Engineers (SPIE) Conference Series},
       volume = {9905},
        month = jul,
          eid = {99052B},
        pages = {99052B},
          doi = {10.1117/12.2235268},
archivePrefix = {arXiv},
       eprint = {1607.00878},
 primaryClass = {astro-ph.IM},
       adsurl = {https://ui.adsabs.harvard.edu/abs/2016SPIE.9905E..2BR}
}

@INPROCEEDINGS{lynx_hdxi,
       author = {{Falcone}, Abraham D. and {Kraft}, Ralph P. and {Bautz}, Marshall W. and {Gaskin}, Jessica A. and {Mulqueen}, John A. and {Swartz}, Doug A.},
        title = "{The high definition x-ray imager (HDXI) instrument on the Lynx X-ray Surveyor}",
    booktitle = {Space Telescopes and Instrumentation 2018: Ultraviolet to Gamma Ray},
         year = 2018,
       editor = {{den Herder}, Jan-Willem A. and {Nikzad}, Shouleh and {Nakazawa}, Kazuhiro},
       series = {Society of Photo-Optical Instrumentation Engineers (SPIE) Conference Series},
       volume = {10699},
        month = jul,
          eid = {1069912},
        pages = {1069912},
          doi = {10.1117/12.2313549},
archivePrefix = {arXiv},
       eprint = {1807.05282},
 primaryClass = {astro-ph.IM},
       adsurl = {https://ui.adsabs.harvard.edu/abs/2018SPIE10699E..12F}
}

@inproceedings{wfi,
author = {Norbert Meidinger and Kirpal Nandra and Markus Plattner},
title = {{Development of the Wide Field Imager instrument for ATHENA}},
volume = {10699},
booktitle = {Space Telescopes and Instrumentation 2018: Ultraviolet to Gamma Ray},
editor = {Jan-Willem A. den Herder and Shouleh Nikzad and Kazuhiro Nakazawa},
organization = {International Society for Optics and Photonics},
publisher = {SPIE},
pages = {312 -- 323},
year = {2018},
doi = {10.1117/12.2310141},
URL = {https://doi.org/10.1117/12.2310141}
}

@article{geant4,
title = "Geant4—a simulation toolkit",
journal = "Nuclear Instruments and Methods in Physics Research Section A: Accelerators, Spectrometers, Detectors and Associated Equipment",
volume = "506",
number = "3",
pages = "250 - 303",
year = "2003",
issn = "0168-9002",
doi = "https://doi.org/10.1016/S0168-9002(03)01368-8",
url = "http://www.sciencedirect.com/science/article/pii/S0168900203013688",
author = "S. Agostinelli and J. Allison and K. Amako and J. Apostolakis and H. Araujo and P. Arce and M. Asai and D. Axen and S. Banerjee and G. Barrand and F. Behner and L. Bellagamba and J. Boudreau and L. Broglia and A. Brunengo and H. Burkhardt and S. Chauvie and J. Chuma and R. Chytracek and G. Cooperman and G. Cosmo and P. Degtyarenko and A. Dell'Acqua and G. Depaola and D. Dietrich and R. Enami and A. Feliciello and C. Ferguson and H. Fesefeldt and G. Folger and F. Foppiano and A. Forti and S. Garelli and S. Giani and R. Giannitrapani and D. Gibin and J.J. [Gómez Cadenas] and I. González and G. [Gracia Abril] and G. Greeniaus and W. Greiner and V. Grichine and A. Grossheim and S. Guatelli and P. Gumplinger and R. Hamatsu and K. Hashimoto and H. Hasui and A. Heikkinen and A. Howard and V. Ivanchenko and A. Johnson and F.W. Jones and J. Kallenbach and N. Kanaya and M. Kawabata and Y. Kawabata and M. Kawaguti and S. Kelner and P. Kent and A. Kimura and T. Kodama and R. Kokoulin and M. Kossov and H. Kurashige and E. Lamanna and T. Lampén and V. Lara and V. Lefebure and F. Lei and M. Liendl and W. Lockman and F. Longo and S. Magni and M. Maire and E. Medernach and K. Minamimoto and P. [Mora de Freitas] and Y. Morita and K. Murakami and M. Nagamatu and R. Nartallo and P. Nieminen and T. Nishimura and K. Ohtsubo and M. Okamura and S. O'Neale and Y. Oohata and K. Paech and J. Perl and A. Pfeiffer and M.G. Pia and F. Ranjard and A. Rybin and S. Sadilov and E. [Di Salvo] and G. Santin and T. Sasaki and N. Savvas and Y. Sawada and S. Scherer and S. Sei and V. Sirotenko and D. Smith and N. Starkov and H. Stoecker and J. Sulkimo and M. Takahata and S. Tanaka and E. Tcherniaev and E. [Safai Tehrani] and M. Tropeano and P. Truscott and H. Uno and L. Urban and P. Urban and M. Verderi and A. Walkden and W. Wander and H. Weber and J.P. Wellisch and T. Wenaus and D.C. Williams and D. Wright and T. Yamada and H. Yoshida and D. Zschiesche"
}

@INPROCEEDINGS{wfi_bkg,
       author = {{von Kienlin}, Andreas and {Eraerds}, Tanja and {Bulbul}, Esra and {Fioretti}, Valentine and {Gastaldello}, Fabio and {Grant}, Catherine E. and {Hall}, David and {Holland}, Andrew and {Keelan}, Jonathan and {Meidinger}, Norbert and {Molendi}, Silvano and {Perinati}, Emanuele and {Rau}, Arne},
        title = "{Evaluation of the ATHENA/WFI instrumental background}",
    booktitle = {Space Telescopes and Instrumentation 2018: Ultraviolet to Gamma Ray},
         year = 2018,
       editor = {{den Herder}, Jan-Willem A. and {Nikzad}, Shouleh and {Nakazawa}, Kazuhiro},
       series = {Society of Photo-Optical Instrumentation Engineers (SPIE) Conference Series},
       volume = {10699},
        month = jul,
          eid = {106991I},
        pages = {106991I},
          doi = {10.1117/12.2311987},
       adsurl = {https://ui.adsabs.harvard.edu/abs/2018SPIE10699E..1IV}
}
\bibliographystyle{spiebib} 

\end{document}